# Iron-based n-type electron-induced

# ferromagnetic semiconductor


Pham Nam Hai[1], Le Duc Anh[1], and Masaaki Tanaka[1,2*]

1. *Department of Electrical Engineering and Information Systems, The University of Tokyo,*

   *7-3-1 Hongo, Bunkyo-ku, Tokyo 113-8656, Japan*

2. *Japan Science and Technology Agency, 4-1-8 Honcho, Kawaguchi-shi, Saitama 332-0012,*

   *Japan*


Carrier-induced ferromagnetic semiconductors (FMSs) have been intensively studied for decades as they have novel functionalities that cannot be achieved with conventional metallic materials. These include the ability to control magnetism by electrical gating[1] or light irradiation[2], while fully inheriting the advantages of semiconductor materials such as band engineering[3]. Prototype FMSs such as (In,Mn)As[4] or (Ga,Mn)As[5-8], however, are always p-type, making it difficult to be used in real spin devices. This is because manganese (Mn) atoms in those materials work as local magnetic moments and acceptors that provide holes for carrier-mediated ferromagnetism. Here we show that by introducing iron (Fe) into InAs, it is possible to



fabricate a new FMS with the ability to control ferromagnetism by both Fe and independent carrier doping. Despite the general belief that the tetrahedral Fe-As bonding is antiferromagnetic, we demonstrate that (In,Fe)As doped with electrons behaves as an n-type electron-induced FMS, a missing piece of semiconductor spintronics for decades. This achievement opens the way to realise novel spin-devices such as spin light-emitting diodes or spin field-effect transistors, as well as helps understand the mechanism of carrier-mediated ferromagnetism in FMSs.



All of semiconductor devices, including pn junction diodes, field effect transistors or semiconductor lasers, require a pair of n-type and p-type semiconductor materials to work. Semiconductor spintronics devices are no exception. Despite the extensive studies on magnetic semiconductors, n-type carrier-induced FMSs are still missing. In fact, most studies on FMSs are concentrated on III-V semiconductor doped with Mn, such as (In,Mn)As[4] or (Ga,Mn)As[5-8], which are always p-type with hole densities as high as $10^{20}$ - $10^{21}$ cm$^{-3}$. In those materials, Mn atoms work not only as local magnetic moments but also as acceptors providing holes that mediate ferromagnetism. This behavior, however, creates a severe drawback; it is difficult to control the ferromagnetism and carrier type (in other words, Fermi level) independently. This problem makes it difficult to utilise the Mn-based FMSs for practical devices, as well as to understand the mechanism of carrier-mediated ferromagnetism in which controlling the Fermi level is very important. On the other hand, II-VI semiconductor based FMSs, such as ZnCrTe[9], are too insulating and there is no effective method for carrier doping. Although there are some reports on enhancing ferromagnetism in ZnCrTe by Iodine (I) doping[10,11], the main effect of this co-doping comes from the enhanced spinodal decomposition of Cr atoms rather than the increase of hole concentration[11]. Indeed, I doped ZnCrTe layers become even more insulating than the undoped ones.

In this paper, we show that by introducing iron (Fe) atoms into InAs, it is possible to fabricate a new FMS with the ability to control ferromagnetism by both Fe and independent



carrier doping. We demonstrate that (In,Fe)As doped with electrons behaves as an n-type electron-induced FMSs, that is, finding the missing counterpart of p-type FMSs.

The studied $(In_{1-x},Fe_x)As$ layers are 100 nm-thick and were grown by low-temperature molecular-beam epitaxy (LT-MBE) on semi-insulating GaAs substrates (see Method summary for sample preparation). Two series of $(In_{1-x},Fe_x)As$ samples were grown as summarized in Table I. Series A with a Fe concentration of $x = 5.0\%$ and series B with a higher Fe concentration of $x = 8.0\%$ (except for B0 with $x = 9.1\%$) were grown at a substrate temperature of 236°C, with and without electron doping. Figure 1a shows a transmission electron microscopy (TEM) image of sample B0, which is undoped $(In_{0.909},Fe_{0.091})As$. Figure 1c shows a high-resolution TEM image of an area close to the buffer layer, indicated by the red rectangular in Fig. 1a. Despite low-temperature growth, the whole (In,Fe)As layer shows zinc-blende crystal structure and no visible inter-metallic precipitation. We further thinned the TEM sample down to ~ 10 nm and found no evidence of such inter-metallic precipitated particles, proving that it is possible to grow zinc-blende (In,Fe)As of good quality by LT-MBE. Figure 1b shows the In, Fe and As atomic concentrations obtained by energy dispersive x-ray (EDX) spectroscopy. It is observed that the As atomic concentration is close to the sum of In and Fe atomic concentrations, revealing that most of the Fe atoms reside at the In sites. The fluctuation of Fe concentration results in superparamagnetic zinc-blende clusters with high Fe concentrations, as will be described later.



At In sites, the Fe ions have two possible states; acceptor state ($Fe^{2+}$) and neutral state ($Fe^{3+}$). If the $Fe^{2+}$ states were dominant, (In,Fe)As layers would be p-type and the hole concentration would be close to the doped Fe concentration at room temperature, similar to the case of (In,Mn)As. In reality, however, sample B0 (and all the other undoped samples) shows n-type with a maximum residual electron concentration of $1.8 \times 10^{18}$ $cm^{-3}$ at room temperature, which is four orders of magnitude smaller than the doped Fe concentration. Our analysis of the temperature dependence of the electron mobility of sample B0 shows that the neutral impurity scattering, rather than the ionized impurity scattering, is the dominant scattering mechanism in this undoped sample up to room temperature (see Fig. S1 in Supplementary Information). All of these facts indicate that the Fe atoms in (In,Fe)As are in the neutral state ($Fe^{3+}$) rather than the acceptor state ($Fe^{2+}$). This result is similar to that obtained in the previous work[12] on paramagnetic (Ga,Fe)As, in which Fe atoms were found to reside at the Ga side and in the $Fe^{3+}$ state. The residual electrons in sample B0 probably come from the As anti-site defects acting as donors due to the LT-MBE growth[13].

We then tried doping (In,Fe)As layers with donors to see the carrier-induced ferromagnetism. After trying several doping methods, we found that Beryllium (Be) atoms doped in (In,Fe)As at a low growth temperature of $T_S = 236°C$ work as good double donors, not as acceptors as in the case of Be-doped InAs grown at $T_S > 400°C$ (see Method summary and Supplementary Information). For these electron doped (In,Fe)As layers, we investigate



their ferromagnetism by using magnetic circular dichroism (MCD), superconducting quantum interference device (SQUID), and anomalous Hall effect (AHE) measurements. Despite the general belief that the tetrahedral Fe-As bonding is antiferromagnetic[14], all of our data show striking evolution of ferromagnetism in (In,Fe)As with increasing both the Fe concentration ($x$ = 5 - 8%) and electron concentration ($n$ = $1.8 \times 10^{18}$ cm$^{-3}$ to $2.7 \times 10^{19}$ cm$^{-3}$), indicating that (In,Fe)As is an intrinsic n-type FMS, and that we can control the ferromagnetism of this material independently by Fe doping and electron doping.

MCD is a technique that measures the difference between the reflectivity of right ($R_{\sigma+}$) and left ($R_{\sigma-}$) circular polarisations: MCD=$\dfrac{90}{\pi}\dfrac{(R_{\sigma+} - R_{\sigma-})}{2} \sim \dfrac{90}{\pi}\dfrac{dR}{dE}\Delta E$, where $R$ is the reflectivity, $E$ is the photon energy, and $\Delta E$ is the spin-splitting energy (Zeeman energy) of a material. Since the MCD spectrum of a FMS directly probes its spin-polarized band structure induced by the s,p-d exchange interactions and its magnitude is proportional to the magnetisation ($\Delta E \sim M$), MCD is a powerful and decisive tool to judge whether a FMS is intrinsic or not[9,15]. Note that the spectral features (i.e. enhanced at optical critical point energies of the host semiconductor), rather than the absolute magnitude of MCD, are important to judge whether a FMS is intrinsic or not. Recently, carrier–induced ferromagnetism was reported in n-type Co-doped TiO$_2$ [16]. However, the intrinsic ferromagnetism in Co-doped TiO$_2$ is controversial, because the MCD spectrum of Co-doped TiO$_2$ does not show enhancement at optical critical point energies of TiO$_2$, while it is



enhanced at energies not related to the band structure of $TiO_2$, and very broad MCD signals are seen at energies smaller than the band gap of $TiO_2$[17]. Figure 2 shows the MCD spectra of sample series A (A1 - A4) and sample series B (B1 - B4), measured at 10 K under a magnetic field of 1 Tesla applied perpendicular to the film plane. With increasing the electron density and Fe concentration, the MCD intensity shows strong enhancement at optical critical point energies $E_1$ (2.61 eV), $E_1 + \Delta_1$ (2.88 eV), $E_0$' (4.39 eV) and $E_2$ (4.74 eV) of InAs, which show the magnetic "fingerprints" of (In,Fe)As. For sample B4, $(In_{0.92},Fe_{0.08})As$ with $n = 2.8 \times 10^{19}$, the MCD peak at $E_1$ already reaches 100 mdeg at 10 K, which is two orders of magnitude larger than the MCD caused by the Zeeman splitting of InAs (~1 mdeg/Tesla)[15]. For a reference, we show in Fig. 2i the MCD spectrum of a 44 nm-thick Fe thin film grown on a GaAs substrate at 30°C. The MCD signals of Fe in the 1.5 – 3.0 eV range are, although quite large, always negative and very broad. Furthermore, there is a very large negative broad peak (-460 mdeg) at around 5.0 eV. In contrast, the MCD signals of (In,Fe)As at 5.0 eV are nearly zero, and there is no broad-spectrum offset background that is the signature of metallic Fe. Furthermore, in (In,Fe)As, the MCD peaks at $E_1 + \Delta_1$ (2.88 eV) and $E_0$' (4.39 eV) are positive, which are consistent with those of (In,Mn)As [15]. All of the above features clearly indicate that the MCD spectra of our (In,Fe)As samples are different from that of Fe, thus eliminating the possibility of metallic Fe particles. These results indicate that (In,Fe)As maintains its zinc-blende structure, and that its spin-split band structure is governed by the s,p-d exchange



interaction between the electron sea and the Fe magnetic moments. Samples A4, B3 and B4, whose electron concentrations are about $10^{19}$ cm$^{-3}$, are ferromagnetic, while other samples with lower electron concentrations are paramagnetic. The facts that the ferromagnetic properties of (In,Fe)As depend on the electron concentration $n$, and that (In,Fe)As can be ferromagnetic only at $n > \sim 10^{19}$ cm$^{-3}$ while paramagnetic at $n < 10^{19}$ cm$^{-3}$, also eliminate the possibility of embedded metallic Fe and intermetallic Fe-As compound particles. At temperatures lower than 236°C, there are three intermetallic Fe-As compounds in their binary phase diagram: FeAs$_2$, FeAs and Fe$_2$As [18]. However, none of them is ferromagnetic; FeAs$_2$ is diamagnetic [19], while FeAs and Fe$_2$As are both anti-ferromagnetic with Neel temperature of 77 K and ~ 353 K, respectively [18].

In the following, we concentrate on the ferromagnetic behaviors of sample A4 and B4. Figures 3a and 3b shows the normalised MCD spectra of sample A4 and B4, measured at 0.2, 0.5 and 1 Tesla. In Fig. 3a, the normalised spectra of sample A4 show nearly perfect overlapping on a single spectrum over the whole photon-energy range, proving that the MCD spectra comes from a single phase ferromagnetism of the whole (In,Fe)As film. In Fig. 3b, the normalised spectra of sample B4 shows perfect overlapping in the range of 2.5 – 5 eV, but deviate slightly from a single spectrum at photon energies lower than 2.5 eV. The peak at 1.8 eV develops faster than that at $E_1$ at low magnetic field, but they approach each other at 1 Tesla. The different behavior between sample A4 and B4 can be more clearly seen by plotting



the normalised MCD intensity as a function of magnetic field (MCD-$H$ curve) at different

photon energies (4.5 eV, 2.6 eV and 1.8 eV), as shown in Figs. 3d and 3e, respectively. While

the magnetic field dependence of MCD of sample A4 measured at different photon energies

perfectly agrees with each other (Fig. 3d), that of sample B4 shows two different behaviors

(Fig. 3e). The MCD intensity at 1.8 eV reaches its saturation value at lower magnetic field

than that at 2.6 and 4.5 eV. This shows that the MCD spectra of sample B4 come from two

ferromagnetic phases. One is the (In,Fe)As matrix phase having a MCD spectrum similar to

that of sample A4 (Fig. 3a), and the other is the cluster phase whose spectrum is enhanced at

low photon energy (< 2.0 eV) as shown in Fig. 3c. The latter turned out to be super

paramagnetic zinc-blende (In,Fe)As clusters with higher density of Fe, as will be clarified

later by SQUID measurements.

Figures 4a and 4b shows the MCD-$H$ curves of sample A4 and B4, respectively,

measured at different temperatures and at the photon energy of 2.6 eV. From these data we

can deduce the Curie temperature of these samples. Because the coercive force and remanent

magnetisation along the perpendicular direction is small due to the shape anisotropy, we use

the Arrott plot technique to estimate the Curie temperature [20]. By using the Arrott plot, we

can also prove the existence of ferromagnetism for the homogeneous sample A4 and the

matrix of sample B4. Figures 4c and 4d show the Arrott plots $MCD^2$ vs. $H$/MCD of sample

A4 and B4 at different temperatures, where MCD is the MCD intensity which is proportional



to the magnetisation $M$. It is clear that sample A4 is ferromagnetic at $T < T_C \sim 34$ K, and the matrix of sample B4 is ferromagnetic at $T < T_{C-1} \sim 28$ K. Next, we confirm these estimations by SQUID measurements. Figures 4e and 4f shows the field cooling (FC) and zero-field cooling (ZFC) magnetisation ($M$) data of sample A4 and B4, measured by SQUID. The magnetic filed is applied in-plane along the GaAs[-110] direction. The $M$-$T$ curves of sample A4 show monotonous behavior both for FC and ZFC, which both rise at $T_C \sim 34$ K, revealing single-phase ferromagnetism. In contrast, sample B4 shows two-phase ferromagnetism. One is the matrix phase with $T_{C-1} \sim 30$ K, and the other is the superparamagnetic phase with $T_{B-2} \sim 35$ K and $T_{C2} \sim 70 \pm 10$ K. Note that the normalised MCD spectrum of sample B4 measured at 50 K (larger than $T_{C-1}$ and $T_{B-2}$ but smaller than $T_{C2}$) still preserves clear features of the zinc-blende InAs structure (Fig. 3c). This fact indicates that these clusters are *not* intermetallic precipitated particles but zinc-blende (In,Fe)As clusters with high concentration of Fe atoms. This is also consistent with the results of microstructure analysis of sample B0 shown in Fig. 1. The formation of the zinc-blende clusters with high concentration of magnetic atoms is the well-known spinodal decomposition phenomena[21,22], which are observed in many FMSs such as (Ga,Mn)As[23,24], ZnCrTe[11] or GeFe[25] with high concentration of magnetic atoms. The $M$-$H$ curves measured with a magnetic field applied along the [-110] direction in the film plane are shown in the inset of Figs. 4e and 4f for sample A4 and B4, respectively. The averaged magnetic moments at saturation are 2.2 and 1.7 $\mu_B$ per doped Fe atom for sample A4 and B4,



respectively. These values are close to that of Mn in (Ga,Mn)As, and surpass most of the theoretically calculated values of the magnetic moment per Fe atom in III-V semiconductors, which are usually smaller than 1 $\mu_B$ [26]. The magnified in-plane $M - H$ curves near the origins (bottom-right of the insets of Figs. 4e and 4f) clearly show hysteresis and remanent magnetisation. The coercive force of sample A4 is 40 Oe. The coercive force of sample B4 is smaller (20 Oe), reflecting the existence of zinc-blende (In,Fe)As clusters. Note that sample B4 does have a shape magnetic anisotropy of the matrix phase, because its $M$-$H$ curve measured with $H$ applied in the film plane saturates much faster than that with $H$ applied perpendicular to the plane. Our study on the in-plane anisotropy in the magnetoresistance of a 10-nm thick n-type $(In_{0.94},Fe_{0.06})As$ layer reveals a two-fold anisotropy along the [-110] direction, and an 8-fold symmetric anisotropy along the crystal axes of (In,Fe)As (see Supplementary Information).

Figures 5a and 5b show the Hall resistance of sample A4 and B4, respectively. The normal Hall effect with negative gradient, showing the n-type conduction of these (In,Fe)As layers, dominates the Hall voltage. The n-type conductivity is also confirmed by the polarity of the thermoelectric Seebeck coefficient, which is $-$ 30 $\mu V/K$ for sample B4 at room temperature (see Fig. S3 in Supplementary Information). There is a small fraction ($\sim$ 3%) of positive anomalous Hall effect (AHE) contribution in both samples due to spin-dependent scattering of electrons at Fe sites, as shown in the anomalous Hall resistance (AHR) curves in



Fig 5c and 5d. The procedure of extracting the AHE component is described in Method summary. The normalised AHE curve of sample A4 perfectly agrees with those of MCD and magnetisation as shown in Fig. 5e, proving again that the ferromagnetism in this sample comes only from the homogenous matrix phase. In contrast, the results of sample B4 are more complicated. In Fig. 5f, the normalised *M-H* curve measured by SQUID lies in the middle of the MCD-*H* curve measured at 1.8 eV (dominated by the superparamagnetic phase) and 2.6 eV (dominated by the matrix phase). This is reasonable since SQUID measures averaged signals from all phases, while MCD can selectively pick up different signals from different phases by changing the photon energy. This fact demonstrates the advantage of the MCD technique in our study. The normalised AHE curve of sample B4 agrees well with the normalised MCD at 2.6 eV at magnetic field smaller than 0.3 Tesla, suggesting that the spin-dependent scattering in the matrix mainly contributes to the AHE at low magnetic field. At magnetic field higher than 0.3 Tesla, the normalised AHE is deviated from the normalised MCD at 2.6 eV, which can be attributed to its non-homogenous structure. For the homogeneous sample A4, we examined its magnetoresistance to find further evidence of spin-dependent scattering. Figure 5g shows the magnetoresistance $\left[\rho(H) - \rho(0)\right]/\rho(0)$ of sample A4 measured at various temperatures, where $\rho(H)$ and $\rho(0)$ are the resistivity at a magnetic field of $H$ and 0, respectively. Clear negative magnetoresistance is observed. The negative magnetoresistance can be understood as the reduction of spin-disorder scattering



when the magnetic moments of Fe atoms are aligned along $H$. Above the Curie temperature (34 K), where the spin-spin correlation between Fe atoms are weak, we can use the spin-disorder scattering formula to describe the magnetoresistance. The spin-disorder scattering resistivity $\rho_s$ is given by [27]

$$\rho_s = 2\pi^2 \frac{k_F}{ne^2} \frac{m^{*2}J_{sd}^2}{h^3} n_{Fe} \left[ S(S+1) - \langle \mathbf{S} \rangle^2 - \langle \mathbf{S} \rangle \tanh\left( \frac{3T_C \langle \mathbf{S} \rangle}{2TS^2(S+1)} \right) \right], \qquad (1)$$

where $k_F$ is the Fermi wave number, $e$ is the elementary charge, $m^*$ is the effective electron mass, $J_{sd}$ is the s-d exchange integral, $h$ is the Plank constant, $n_{Fe}$ is the density of Fe, $S$ is the Fe spin, and $\langle \mathbf{S} \rangle$ is the thermal average of $S$. For small magnetic fields, $\langle \mathbf{S} \rangle \ll S$ and $\tanh\left( \frac{3T_C \langle \mathbf{S} \rangle}{2TS^2(S+1)} \right) \simeq \frac{3T_C \langle \mathbf{S} \rangle}{2TS^2(S+1)}$; thus, the magnetoresistance ratio $\left\| [\rho(H) - \rho(0)] / \rho(0) \right\|$ is simply proportional to $M^2$. Since AHR is proportional to $M$ as evidenced in Fig. 5e, a linear relationship between $\left\| [\rho(H) - \rho(0)] / \rho(0) \right\|$ and AHR$^2$ should be expected. Figure 5h shows $\left\| [\rho(H) - \rho(0)] / \rho(0) \right\|$ vs. AHR$^2$ plotted at $T$ = 40 – 70 K. Excellent linear relationships between $\left\| [\rho(H) - \rho(0)] / \rho(0) \right\|$ and AHR$^2$ are observed, indicating that both the observed AHE and negative magnetoresistance originate from the spin-dependent scattering in this (In,Fe)As sample, and that parallel conduction in the buffer layer is negligible. Note that the observed negative magnetoresistance is different from those observed in systems with ferromagnetic nanoclusters or weak localization (see Supplementary Information).

In Figures 6a and 6b, we show the evolution of ferromagnetism expressed by $T_C$ vs.



electron concentration and resistivity vs. temperature of series A and B samples, respectively. It is clear that there is a threshold electron concentration of about $10^{19}$ cm$^{-3}$ for (In,Fe)As to become ferromagnetic. The steep change in magnetic behavior at $10^{19}$ cm$^{-3}$ shown Fig. 6a is clearly correlated with the metal-insulator transition of (In,Fe)As layers as shown in Fig. 6b. All of these results confirm that (In,Fe)As is an intrinsic n-type ferromagnetic semiconductor whose ferromagnetism is induced by electrons. It should be noted that the homogenous sample A4 with $T_C$ as high as 40 K requires only an electron concentration of $1.8 \times 10^{19}$ cm$^{-3}$. Comparing with (In,Mn)As, this electron concentration is an order of magnitude smaller ($T_C \sim$ 20 K requires 1.0 - $1.6 \times 10^{20}$ cm$^{-3}$ of holes for (In,Mn)As, see Ref. 1). Noting that a carrier concentration change of $\sim 10^{20}$ cm$^{-3}$ can be obtained by applying a gate voltage in field-effect transistor structures[37], this small electron concentration gives (In,Fe)As another advantage over (In,Mn)As when controlling ferromagnetism by electrical and optical means.

What can be expected using an n-type FMS? There have been already a large number of proposed spin-devices using pn junctions with a p-type FMS and non-magnetic n-type semiconductor (SC) or vice-versa, in which carrier spins in non-magnetic layers are generated by irradiating circularly polarized light [28-31]. With n-type (In,Fe)As, we can realize spin-devices with much enhanced performance without any external light source. In the simplest case of an all FMS magnetic pn junction, one can fabricate a spin-diode structure whose forward current can be modulated by changing the relative magnetisation direction of



the p(n) layers in a fashion similar to that of the giant magnetoresistance effect [28]. A spin-diode can be used as a magnetic field sensor or non-volatile memory, which can be fully integrated in a semiconductor electronic circuit. If the recombination rate of electron-hole pairs in the depletion layer is large, we can obtain an all FMS spin-LED emitting circularly polarized light. A natural extension of ferromagnetic pn junctions is an all-FMS bipolar transistors, with which we can modulate the collector current by changing the relative direction of magnetisation of the FMS layers [31, 32]. Spin metal-oxide-semiconductor field-effect transistors (spin MOSFETs) with n-type FMS source, drain and p-type FMS channel (or vise verse)[33] can be used for high-density magnetic memory [34] or reconfigurable logic circuits [35]. In such spin MOSFETs, we can fully utilize the carrier-induced ferromagnetism of FMSs to control the ferromagnetic behavior of the channel, such as minimizing the magnetisation for smaller switching field of the channel magnetisation [1,36,37].

Using Fe as magnetic dopants has another important advantage over Mn, especially when studying the mechanism of carrier-induced ferromagnetism. In the case of (Ga,Mn)As, there are Mn-related impurity states, which complicate the theory of carrier-induced ferromagnetism. In contrast, Fe atoms in III-V are neither major donors nor acceptors; thus, there are probably no available Fe-related donor or acceptor impurity states. The original mean-field Zener model of carrier-induced ferromagnetism in Mn based FMSs was developed based on the assumption that holes reside in the valence band (VB). For example, the Fermi



level of (Ga,Mn)As is assumed to be 200-300 meV below the top of the VB (the VB model) [38,39]. The ferromagnetism in this model was explained by the *p-d* exchange interaction between the VB holes and the localized Mn-3d electrons. Until recently, this model has been widely accepted as the standard theory of carried-induced ferromagnetism in Mn based FMS, since it can explain some features of (Ga,Mn)As [37,40,41]. On the other hand, recent reports on the optical [42-44] and transport [45-47] properties of (Ga,Mn)As have shown that holes exist in the impurity states within the band gap of (Ga,Mn)As with an effective mass as heavy as $10m_0$, where $m_0$ is the free electron mass. Those results make the assumption of mean-field Zener model unjustified, and suggest an alternative model called the impurity band (IB) model. Although the debate on the band structure of (Ga,Mn)As is still in progress, it is clear that the difficulty in understanding the ferromagnetism in (Ga,Mn)As comes from the existence of such IB in the band gap, with which it is difficult to deal theoretically and experimentally. For example, the MCD spectra calculated for the VB and IB models of (Ga,Mn)As by the first principles calculation show no clear difference [48]. Because the Fermi level in the IB in (Ga,Mn)As is just several tens of meV above the top of the VB, it is not possible to distinguish whether the $E_0$(1.6 eV) peak in MCD spectra comes from the optical transition of holes in the VB or from those in the IB. As a result, strong MCD $E_0$ peak does not immediately mean the large spin-splitting of the valence band in (Ga,Mn)As. Careful resonant tunneling experiments through (Ga,Mn)As quantum wells have recently shown that the VB of



(Ga,Mn)As is nearly non-magnetic [47]. In contrast to the results of (Ga,Mn)As, the MCD peaks at $E_1$ (2.61 eV), $E_1 + \Delta_1$ (2.88 eV), $E_0$' (4.39 eV) and $E_2$ (4.74 eV) in our (In,Fe)As samples (see Fig. 2) come from optical transitions that are not related to energy levels near the Fermi energy of the conduction band. Therefore, the corresponding band edges of InAs should be spin-split. This argument is supported by the fact that the positions of these peaks are independent of the Fe and electron density, while the MCD peak $E_0$ (1.6 eV) of (Ga,Mn)As can shift by 0.2 eV when changing the Mn and hole concentrations [49].

In the following, we show that electrons in (In,Fe)As are in the conduction band of InAs, and not related to any Fe hypothetical d-band or itinerant impurity states. This greatly reduces the complexity of interpretation of the ferromagnetism in this material. To investigate the band structure where electrons reside in (In,Fe)As, we have measured the thermoelectric Seebeck coefficient of our samples at room temperature. The measurement setup and result are given in Supplementary Information. Because all samples clearly show negative Seebeck coefficients, it was confirmed that the conduction carriers in (In,Fe)As are electrons. Furthermore, from the electron density and the Seebeck coefficient, we can estimate the electron effective mass, which is $0.030 \sim 0.171 m_0$ depending on the electron concentrations. These data are all consistent with the effective mass of electrons reported in heavily doped InAs[50], and indicate that the electrons in (In,Fe)As reside in the conduction band with a light effective mass rather than in the hypothetical Fe-related itinerant impurity band with a heavy



effective mass. The Fermi energy $E_F$ measured from the bottom of the conduction band is 0.14 ~ 0.36 eV. All of these features make (In,Fe)As a very ideal FMS material.

Using $m^* = 0.171 m_0$ and $E_F = 146$ meV for sample A4, we can roughly estimate the s-d exchange interaction, expressed in terms of $\left|N_0 J_{sd}\right|$, where $N_0$ is the density of cation sites. The maximum reduction of resistivity due to the suppression of spin disorder scattering is given by $\Delta\rho_{\max} = 2\pi^2 \dfrac{k_F}{ne^2} \dfrac{m^{*2} J_{sd}^2}{h^3} n_{Fe} S(S+1)$, which can be roughly estimated by $\rho(0)$ -$\rho(H_{\max})$ when $T$ approaches 0 K, where $H_{\max}$ is the maximum applied magnetic field. Using the magnetoresistance data measured at 3.5 K ($\Delta\rho_{\max} = 5.7\times10^{-5}$ $\Omega$cm), we roughly estimated $\left|N_0 J_{sd}\right| \sim 1.1$ eV. This value is quite reasonable comparing with those of II-VI diluted magnetic semiconductors.

In conclusion, we have grown a Fe-based n-type electron-induced FMS, (In,Fe)As. MCD, SQUID, and magnetotransport data show clear evolution of ferromagnetism in (In,Fe)As when increasing the electron density by chemical doping with a fixed Fe concentration. The normal Hall effect and thermoelectric Seebeck effect confirm the n-type conduction of (In,Fe)As. The effective mass data show that electrons reside in the conduction band, not in the hypothetical Fe-related impurity band. Development of such n-type Fe-based FMS will open the way to fabricate all-FMS spintronic devices, as well as help understanding the physics of carrier-induced ferromagnetisms in FMS.



**Method summary**

All samples were grown by molecular beam epitaxy on semi-insulating GaAs substrates. After growing a 50 nm-thick GaAs buffer layer at 580°C, we grew a 10 ~ 20 nm-thick InAs buffer layer at 500°C. The growth of InAs at high temperature helps relax quickly the lattice mismatch between InAs and GaAs, and create a relatively smooth InAs surface. After cooling the sample down to 236°C, we started growing a 100 nm–thick (In,Fe)As with or without Be co-doping. Finally, we grew a 5 ~ 10 nm InAs cap (except for sample B0 with a 20 nm cap) to prevent oxidation of the underlying (In,Fe)As layer. At this low growth temperature, we found that doped Be atoms act as donors rather than acceptors (see Supplementary Information). As-grown samples were cleaved to small pieces for MCD and Hall effect measurements without any treatment. The Hall effect was measured in the Van der Pauw configuration. To eliminate the effect of the magnetoresistance due to misalignment of the Hall voltage terminals, we took the odd function from the raw data. To extract the normal and anomalous Hall effect component, we subtract from the original Hall effect data a linear component (normal Hall effect $\sim \frac{1}{ne}$), so that the remaining non-linear component (AHE $\sim M$) has the same zero-field susceptibility as that obtained by SQUID magnetometry or MCD. To prepare samples for SQUID measurements, we coated the surface with paraffin, then dipped the samples to HCl acid solution for 1 hour to remove any possible magnetic impurities attached to the substrate. To avoid any remanent magnetic field in our superconducting magnet, by using the thermal switch of the magnet, we confirmed that there is no remaining current in the superconducting coil at zero magnetic field in every *M-H* scan.

**Acknowledgements** This work was partly supported by Grant-in-Aids for Scientific Research, the Special Coordination Programs for Promoting Science and Technology, the FIRST Program of JSPS, and the Global COE program (C04). The authors acknowledge Dr. Fuji for his help in the measurement of Seebeck effect. Part of this work was done at the Cryogenic Center, the Univ. Tokyo.

**Author contributions** P. N. H. designed the experiment and fabricated the samples. P. N. H and L.D.A collected data and performed analysis of those data; M.T. managed and planned the research and supervised the experiment. All authors discussed the results and commented on the manuscript.

**Figure Legends**

**Figure 1. Microstructure analysis.** **a,** Transmission electron microscopy (TEM) image of a 100 nm-thick $(In_{0.909},Fe_{0.091})As$ layer (sample B0 in table I) grown on a GaAs substrate, taken from the GaAs[110] direction. **b,** In, Fe and As atomic concentrations obtained by energy dispersive X-ray spectroscopy (EDX) taken at 6 points marked by * in the above TEM image. It is observed that the As atomic concentration (~50%) is close to the sum of the In and Fe atomic concentrations, revealing that Fe mostly reside at the In site, although there are fluctuations of Fe concentration depending on the location. **c,** High-resolution TEM (HRTEM) lattice-image taken at an $(In_{0.92},Fe_{0.08})As$ area close to the substrate (marked by the



red rectangular in Fig. 1a). The (In,Fe)As lattice shows zinc-blende crystal structure only. Other HRTEM images taken at areas close to the surface and in the middle of this (In,Fe)As layer show no inter-metallic precipitation, although there are some stacking faults due to the large lattice mismatch between the (In,Fe)As and GaAs substrate.

**Figure 2. Magnetic circular dichroism spectra (MCD) of   a-d, (**$In_{0.95}Fe_{0.05}$**)As samples** (A1 - A4 in table I) with electron concentrations of $1.8\times10^{18}$, $2.9\times10^{18}$, $6.2\times10^{18}$, $1.8\times10^{19}$ cm$^{-3}$, respectively, measured at 10 K and under a magnetic field of 1 Tesla applied perpendicular to the film plane, and   **e-h, (**$In_{0.92}Fe_{0.08}$**)As samples** (B1 - B4 in table I) with electron concentrations of $1.3\times10^{18}$, $1.5\times10^{18}$, $9.4\times10^{18}$, $2.8\times10^{19}$ cm$^{-3}$, respectively. With increasing the electron and Fe concentrations, the MCD spectra show strong enhancement at optical critical point energies $E_1$ (2.61 eV), $E_1 + \Delta_1$ (2.88 eV), $E_0$' (4.39 eV) and $E_2$ (4.74 eV) of InAs. **i,** MCD spectrum of a 44 nm-thick Fe thin film grown on a GaAs substrate at 30°C. The spectrum is clearly different from those of (In,Fe)As.

**Figure 3. Normalised MCD spectra and hysteresis.   a,** Normalised MCD spectra of sample A4 with $H$ = 0.2, 0.5 and 1 Tesla, measured at 10 K. A single spectrum over the whole photon-energy range proves the single phase ferromagnetism of this sample.   **b-c,** Normalised MCD spectra of sample B4 with $H$ = 0.2, 0.5 and 1 Tesla, measured at 20 and 50



K, respectively. The spectra at 20 K shows two phase ferromagnetism at low photon energy (< 2.0 eV), and can be decomposed to two components. One is the matrix with spectrum similar to the sample A4 (Fig. 3a), and the other is the cluster phase whose spectrum is enhanced at low photon energy (< 2.0 eV) as shown in Fig. 3c. **d-e,** Normalised MCD – magnetic field (MCD - $H$) curves of samples A4 and B4, respectively, measured at photon energies of 1.8, 2.6 and 4.5 eV. The MCD - $H$ curves of sample A4 perfectly coincide with each other, while that of sample B4 at 1.8 eV shows smaller saturation field than that at 2.6 and 4.5 eV.

**Figure 4. Temperature dependence. a,b,** MCD-$H$ curves at different temperatures of sample A4 and B4, respectively. These curves are measured at the photon energy of 2.6 eV. **c,d,** Arrott plot (MCD$^2$ vs. $H$/MCD) of sample A4 and B4. Sample A4 is ferromagnetic at $T < T_C \sim$ 34 K. The matrix of sample B4 is ferromagnetic at $T < T_{C-1} \sim$ 28 K. **e,f,** magnetisation ($M$ - $T$ curves) of sample A4 and B4, measured under 1-Tesla field-cooling (FC) and zero-field-cooling (ZFC) conditions. The magnetic field (20 Oe) is applied in-plane along the GaAs[-110] direction. Sample A4 shows a single-phase ferromagnetism with Curie temperature $T_C \sim$ 34 K. In contrast, sample B4 shows two-phase ferromagnetism: One is the matrix phase with $T_{C-1} \sim$ 28 K, and the other is the cluster phase with blocking temperature $T_{B-2} \sim$ 35 K and Curie temperature $T_{C-2} \sim$ 70±10 K. The insets show the magnetisation hysteresis loops ($M$ - $H$) of sample A4 and B4 measured at 10 K. The magnified $M$ - $H$ curves



near the origin are shown in the bottom-right of the insets, which clearly show the remanent magnetisation.

**Figure 5. Transport characteristics.** **a-b,** Hall resistances of sample A4 and B4, respectively. The Hall resistance is dominated by the normal Hall effect with negative gradient, showing the n-type conduction of these samples. **c-d,** The extracted *positive* anomalous Hall resistances (AHR) for sample A4 and B4, respectively. The AHR are about 3% of the normal Hall resistances, even at 10 K. Nevertheless, clear temperature dependence of these AHR was observed. **e-f,** Comparison of the magnetic field dependence of MCD, magnetisation and AHR at 10 K for sample A4 and B4, respectively. **g,** magnetoresistance of sample A4, normalised by its value at zero magnetic field. The magnetic field is applied perpendicular to the film plane. **h,** magnetoresistance ratio $\left\| [\rho(H) - \rho(0)] / \rho(0) \right\|$ vs. $AHR^2$ of sample A4 measured at $40 - 70$ K. Excellent linear relationships indicate that both AHE and magnetoresistance originate from the spin-dependent scattering in the (In,Fe)As layer, not from the parallel conduction.

**Figure 6. a,** $T_C$ vs. electron concentration and **b,** resistivity vs. temperature summarized for sample series A and B. An electron concentration threshold of about $10^{19}$ cm$^{-3}$ is needed for ferromagnetism, which is also the boundary for metal-insulator transition.



**Table I.  List of (In$_{1-x}$,Fe$_x$)As samples. All samples were grown at 236°C.**

| Sample | Fe concentration $x$ (%) | Electron concentration $n$ (cm$^{-3}$) | Non-magnetic dopants |
|--------|--------------------------|----------------------------------------|----------------------|
| A1 | 5.0 | $1.8 \times 10^{18}$ | Be |
| A2 | 5.0 | $2.9 \times 10^{18}$ | Be |
| A3 | 5.0 | $6.2 \times 10^{18}$ | Be |
| A4 | 5.0 | $1.8 \times 10^{19}$ | Be |
| B0 | 9.1 | $1.6 \times 10^{18}$ | None |
| B1 | 8.0 | $1.3 \times 10^{18}$ | Be |
| B2 | 8.0 | $1.5 \times 10^{18}$ | Be |
| B3 | 8.0 | $9.4 \times 10^{18}$ | Be |
| B4 | 8.0 | $2.8 \times 10^{19}$ | Be |



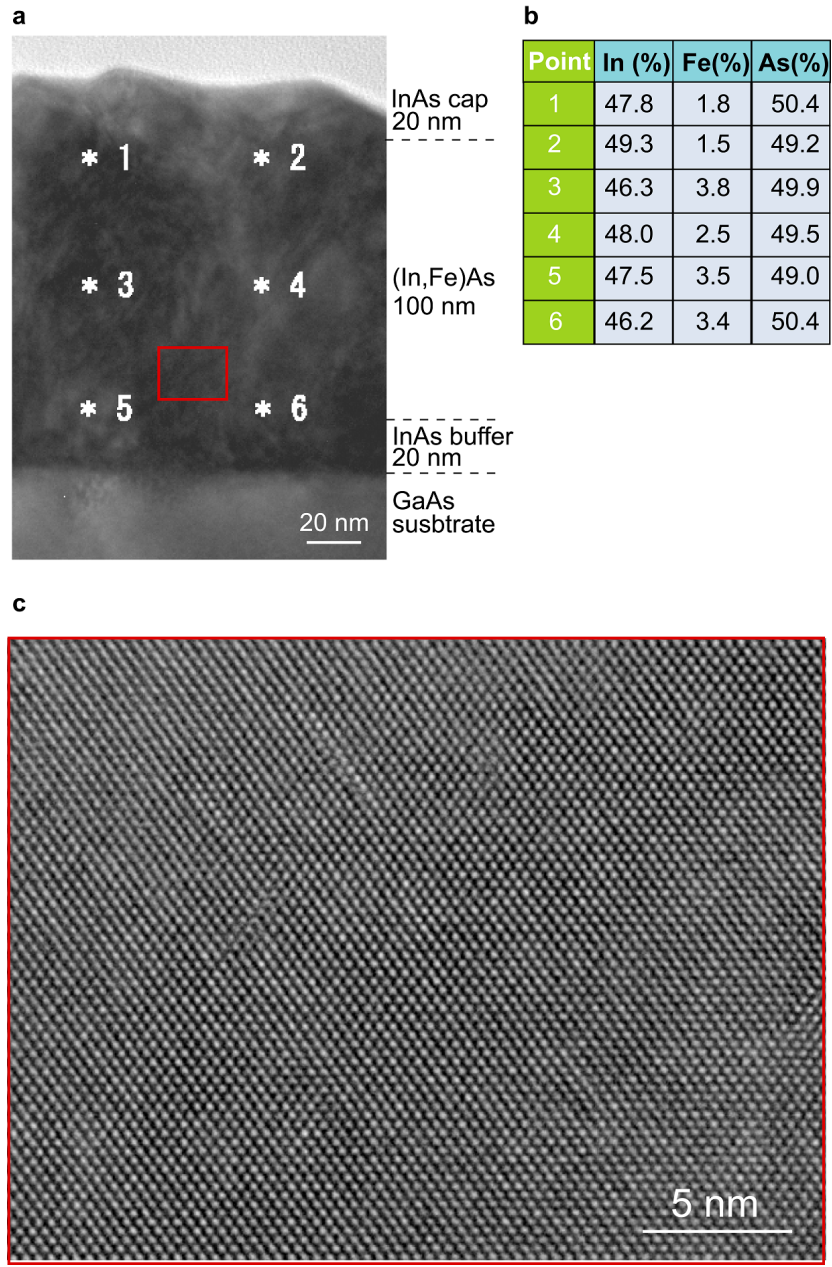

Fig. 1. Hai *et al.*



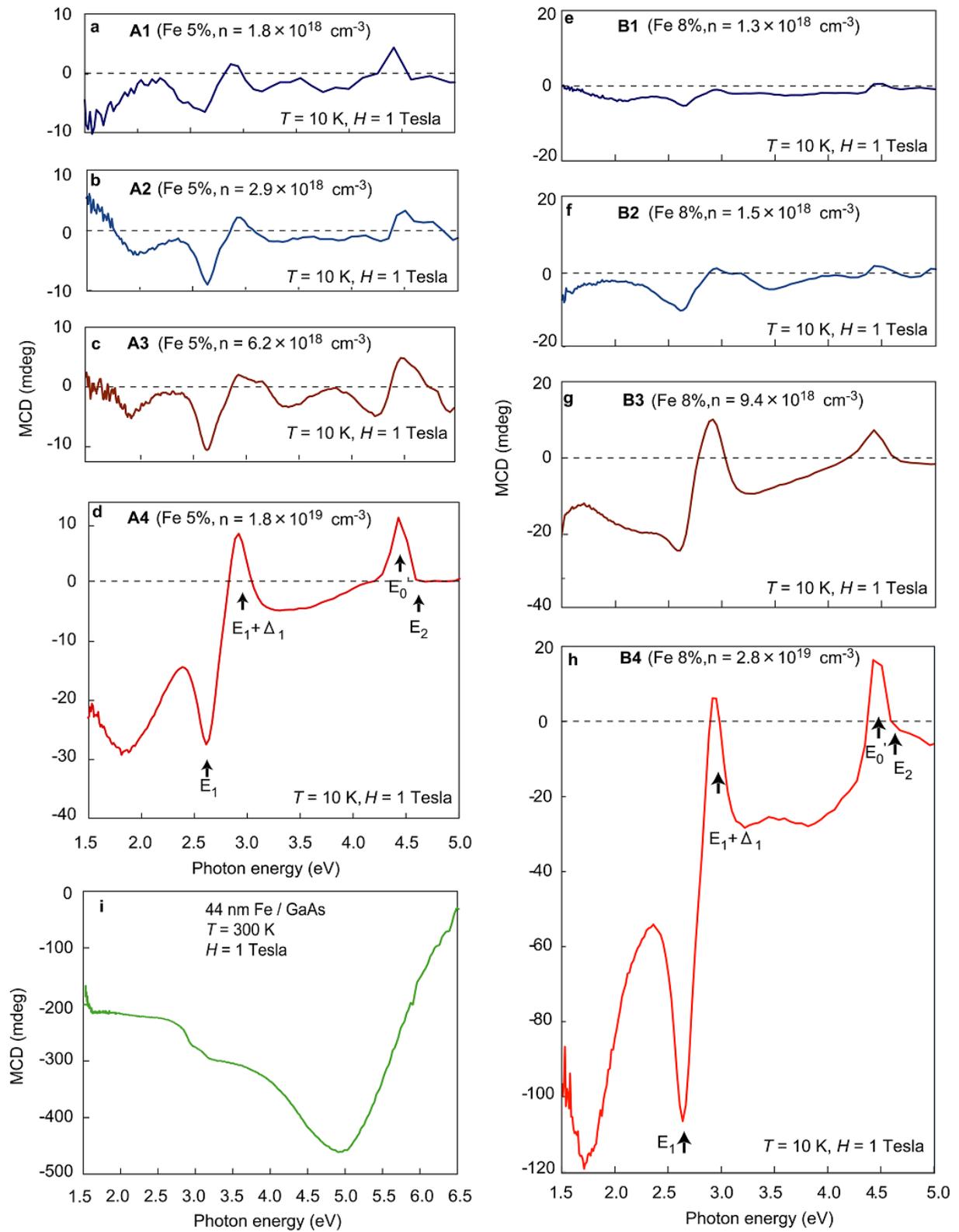

Fig. 2. Hai *et al.*



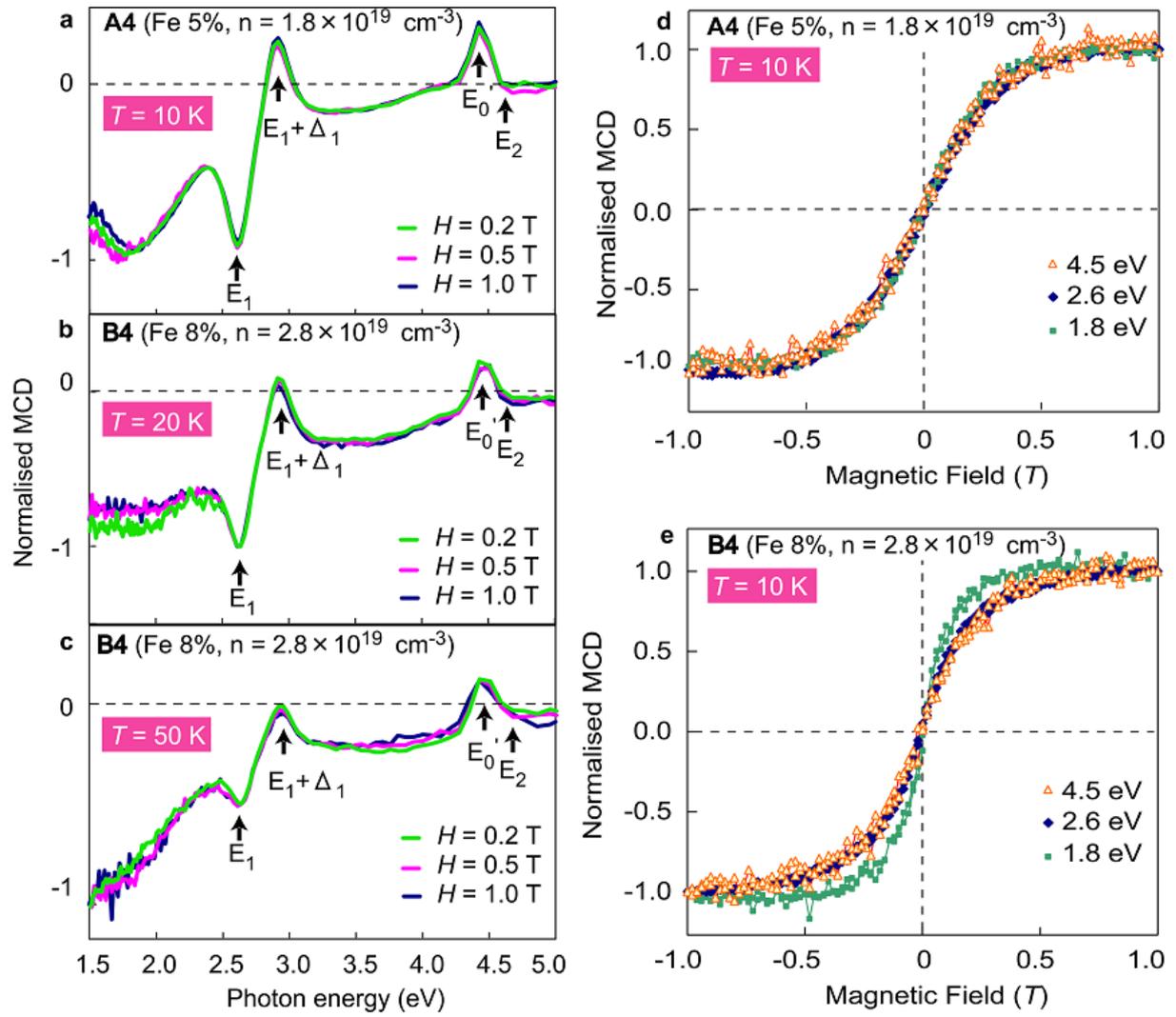

Fig. 3. Hai *et al.*



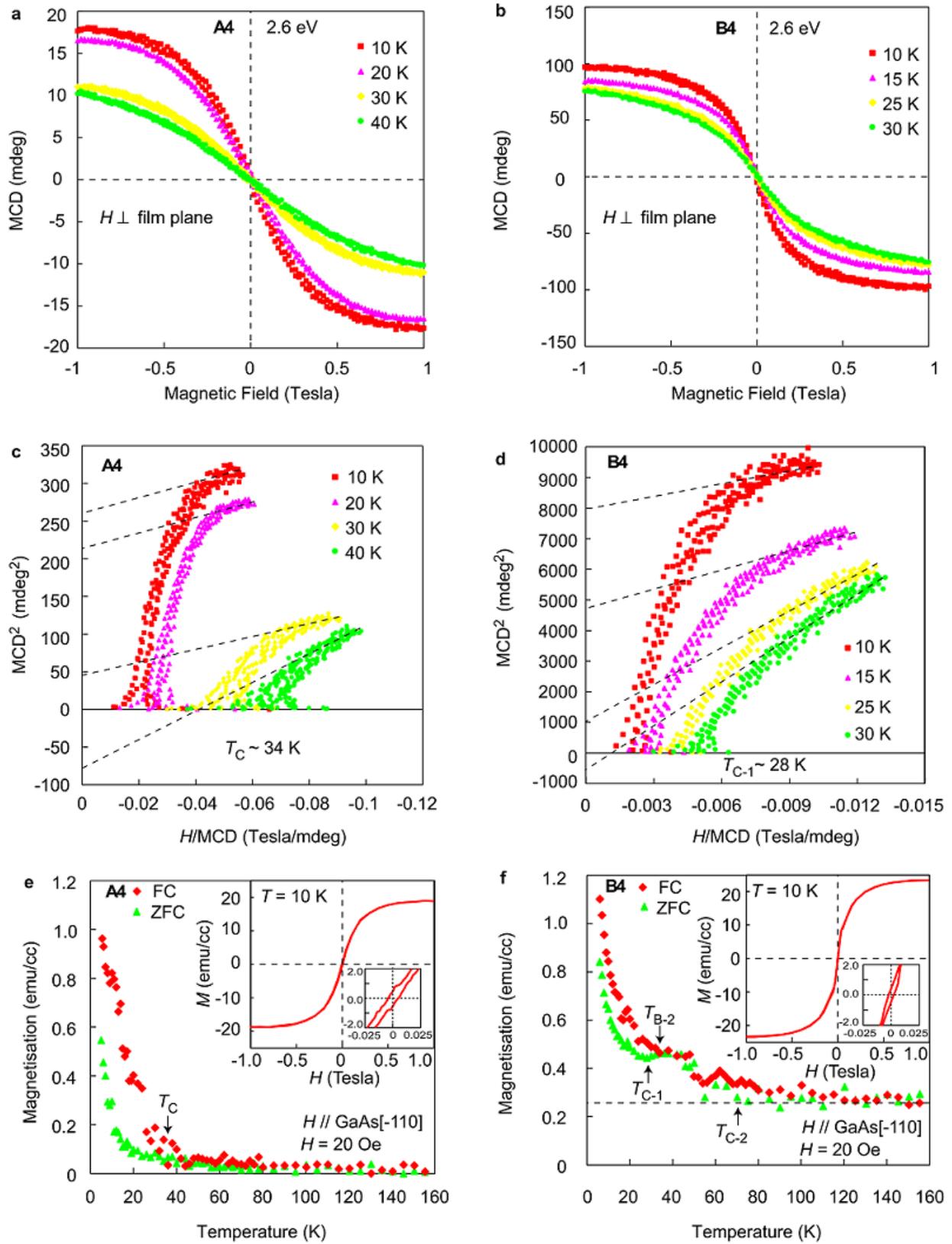

Fig. 4. Hai *et al.*



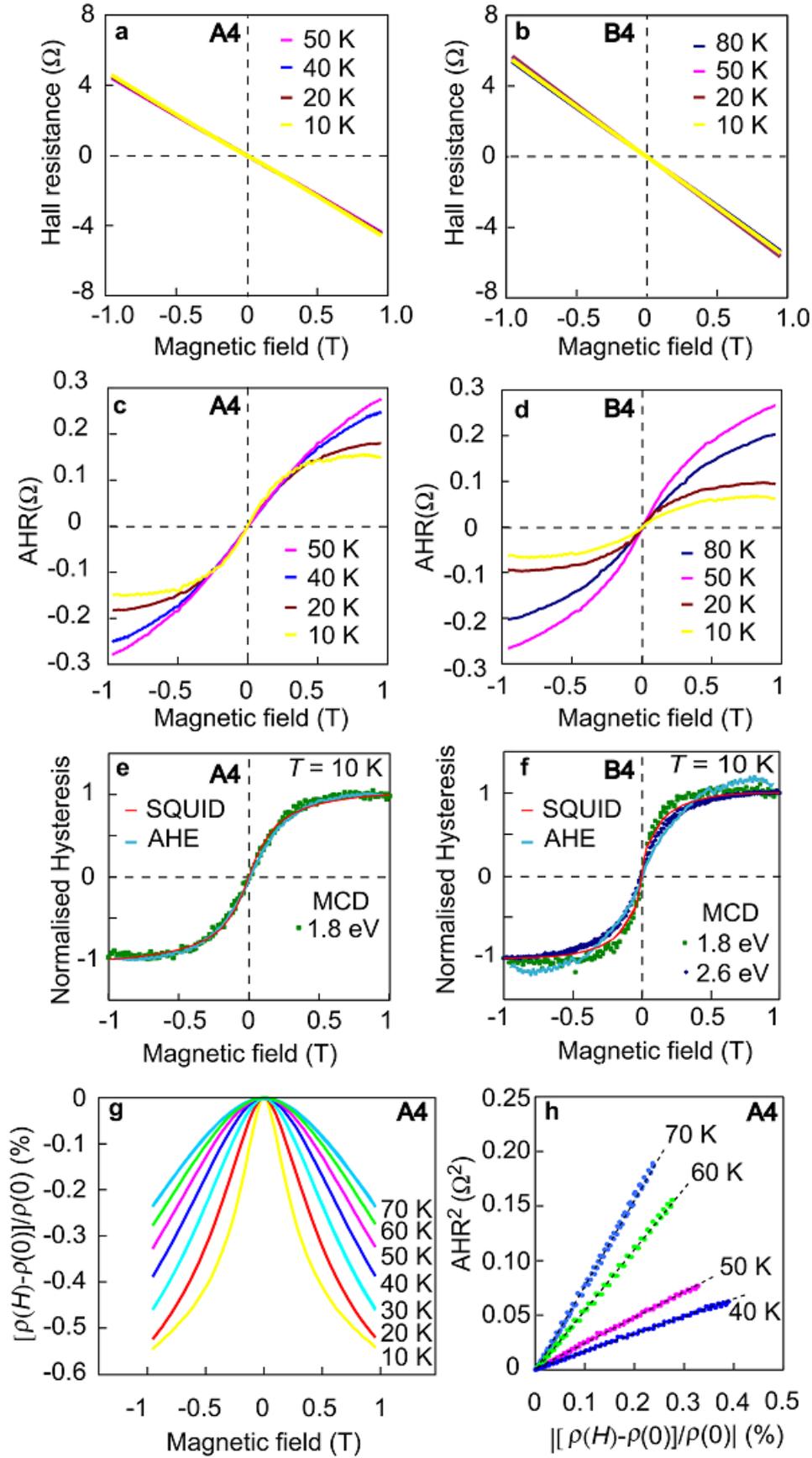

Fig. 5. Hai *et al.*



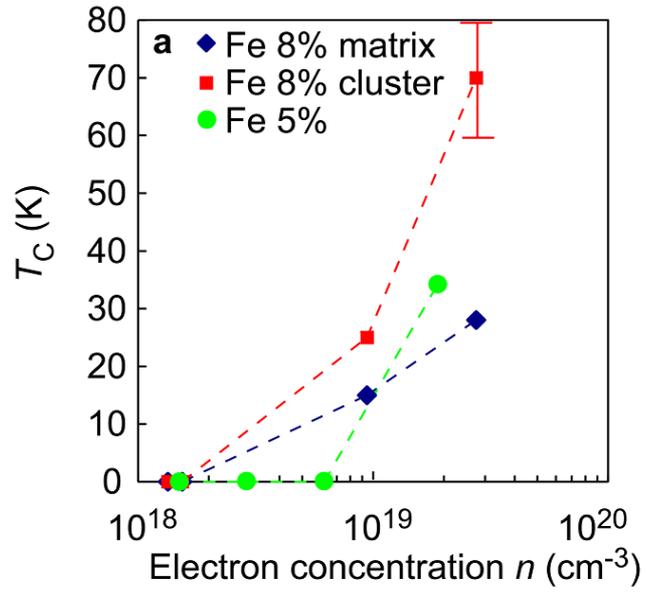

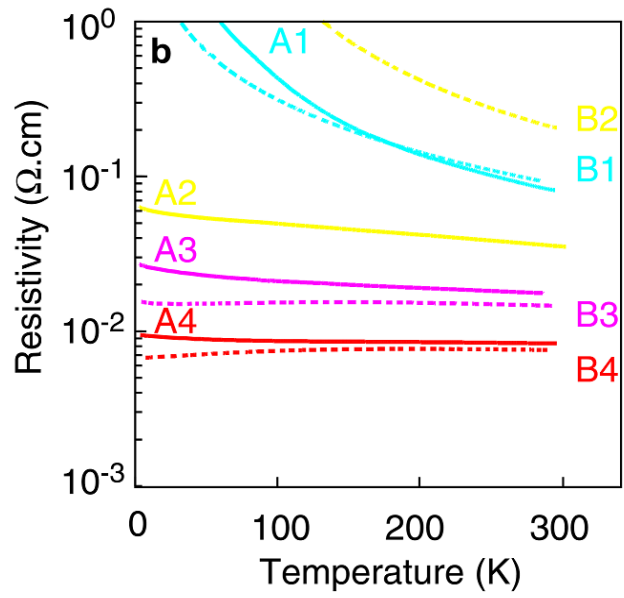

Fig. 6. Hai *et al.*





# Iron-based n-type electron-induced ferromagnetic semiconductor

Pham Nam Hai, Le Duc Anh, Masaaki Tanaka

**Temperature dependence of the mobility of sample B0**

Fig. S1 shows the temperature dependence of the electron mobility $\mu$ of the as-grown $(In_{0.909},Fe_{0.091})As$ (sample B0), with vertical and horizontal axes plotted in the logarithmic scale. The dashed red line is the fitting $\mu \sim T^{\gamma}$. For $T < 50$ K, the mobility is nearly temperature-independent ($\gamma = 0.04$). For $T > 50$ K, the mobility still weakly depends on temperature ($\gamma = 0.25$). These suggest that the Fe impurities in this material remain neutral. If the Fe impurities were ionized (i.e. in the acceptor $Fe^{2+}$ state), the sample would be p-type and the $\mu - T$ relation would be given by $\mu \sim \dfrac{1}{n_{ion}}(2k_B T)^{3/2}[\ln(1+\alpha k_B^2 T^2)]$ for ionized impurity scattering, requiring $\gamma \geq 1.5$. In reality, the sample is n-type and $\gamma$ is close to zero, indicating that the Fe impurities remain in the neutral state. When the neutral impurity scattering dominates, the $\mu - T$ relation is given by $\mu \sim \dfrac{1}{n_{neutral}} \times const$, which is nearly temperature-independent. Therefore, the Fe impurities on In sites should be in the $Fe^{3+}$ state. This result is similar to that obtained for paramagnetic (Ga,Fe)As [1]. Since the Fe impurities contribute to spin but not to carrier generation, we have an important degree of freedom for controlling the carrier type and carrier concentration by independent chemical doping.

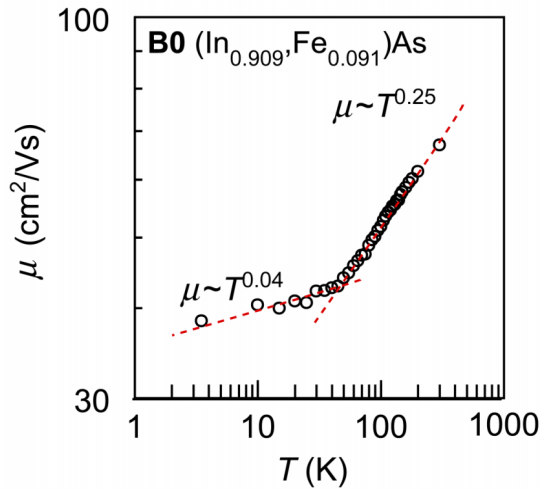

**Fig. S1.** Temperature dependence of the mobility of sample B0, $(In_{0.909},Fe_{0.091})As$, which indicates the neutral state of Fe impurities on In sites. Dashed red line is the fitting $\mu \sim T^{\gamma}$.



**Thermoelectric Seebeck effect at room temperature.**

A convenient way to confirm the carrier type and measure the effective mass in heavily doped semiconductors is the thermoelectric Seebeck effect. Figure S2a shows the principle of the Seebeck effect. When there is a temperature gradient $\Delta T$ between two edges of a sample, carriers at the hot side are more thermally activated and then diffuse to the cold side until equilibrium is established. As a result, a voltage $\Delta V$ will be generated between the edges. The Seebeck coefficient $\alpha$ of a material is defined as $\alpha = -\Delta V/\Delta T$. If carriers are electrons, $\alpha$ is negative. Inversely, if carriers are holes, $\alpha$ is positive. Figure S2b shows the experimental setup to measure the Seebeck effect of our (In,Fe)As at room temperature. The hot side is a copper (Cu) electrode with a heater, placed on an epoxy film. The epoxy film acts as a thermal insulator. The cold side is a Cu electrode placed on a sapphire substrate, which acts as a thermal sink. A piece of sample bridges the hot and cold electrodes. Silver paste is used for electrical contacts between the edges of the sample and the electrodes. Voltage signals from a thermocouple made from Cu wire (red thin line) and Constantan (green thin line) measure the temperature difference $\Delta T_{raw}$ between the hot Cu electrode and the sapphire substrate when the heater is turned on. Figure S2c shows the measured $\Delta V$ - $\Delta T_{raw}$ of sample B4 at different heater currents. It is clear that $\alpha$ is negative from the gradient of this data. Thus, the carriers are electrons, which is consistent with the Hall effect measurement results described in the main text.

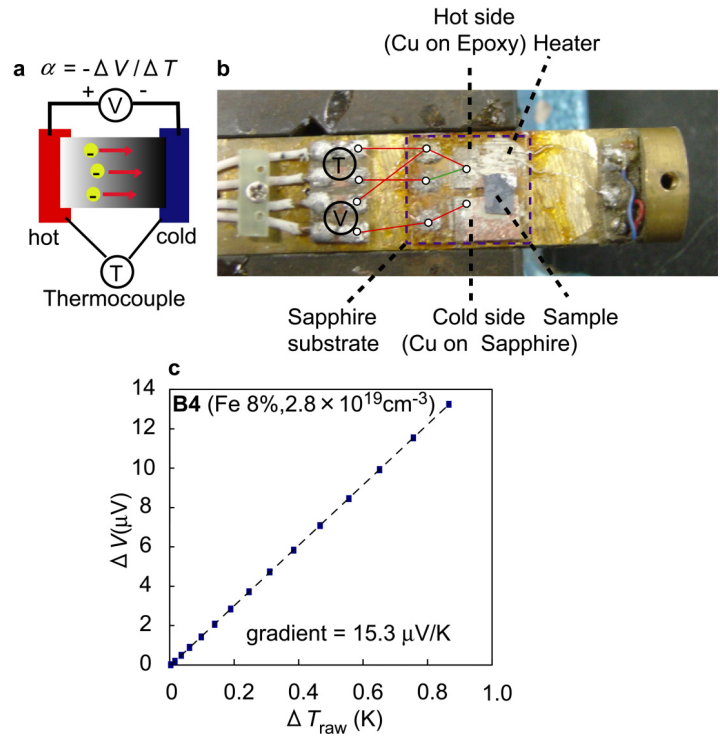

**Fig. S2. a,** Principle of thermoelectric Seebeck effect. **b,** Experiment setup to measure the Seebeck effect. **c,** Measured $\Delta V$ - $\Delta T_{raw}$ of sample B4.



Using the values of $\alpha$ and electron concentration $n$, we can estimate the effective mass $m^*$ of electrons and the Fermi energy $E_F$ by solving the following equations:

$$\alpha = -\frac{k_B}{e}\frac{\pi^2}{3}\left(s+\frac{3}{2}\right)\frac{k_B T}{E_F}, \tag{S1}$$

$$n = \frac{4N_C}{3\sqrt{\pi}}\left(\frac{E_F}{k_B T}\right)^{3/2}, \tag{S2}$$

$$N_C = 2\left(\frac{m^* k_B T}{2\pi\hbar^2}\right)^{3/2}. \tag{S3}$$

Here $k_B$ is the Boltzmann constant, $e$ is the elementary charge, $N_C$ is the effective density of state. $s$ in Eq. (S1) is the exponent of the scattering time $\tau \sim \varepsilon^{-s}$. Here we use $s = 0$ for neutral impurity scattering.

The electron concentration $n$ can be easily obtained from the Hall effect measurement at room temperature. Note that the anomalous Hall effect is quite small compared with the normal Hall effect even at low temperature, so we can neglect its contribution at room temperature. The magnitude of $\alpha$ is given by $\alpha = -\frac{\Delta V}{\Delta T_{edge}} = -\left(\frac{\Delta T_{raw}}{\Delta T_{edge}}\right)\frac{\Delta V}{\Delta T_{raw}} = -k\frac{\Delta V}{\Delta T_{raw}}$,

where $k$ is the ratio between the measured $\Delta T_{raw}$ and the real temperature difference $\Delta T_{edge}$ between the two edges of the sample. If the thermal conductivity of a sample is much smaller than those of copper and sapphire, then $\Delta T_{raw} = \Delta T_{edge}$. In reality, due to the good thermal conductivity of GaAs, there is a temperature distribution in the electrodes and sapphire substrate. As a result, $\Delta T_{edge}$ is generally smaller than $\Delta T_{raw}$. $k$ is measured to be 2 for a reference sapphire sample, whose thermal conductivity $\sim 0.42$ W/(cm·degree) is nearly equal to 0.44 W/(cm·degree) of semi-insulating GaAs. Therefore, in this experiment, we multiply the gradient of $\Delta V$ - $\Delta T_{raw}$ data by $-k = -2$ to obtain the magnitude of $\alpha$. For example, $\alpha$ of sample B4 is estimated to be -30 μV/K from the data of Fig. 2Sc.

Color plots in Figure S3 show the obtained effective mass $m^*$ of our several (In,Fe)As samples (series A and B in this work) with varying the Fe concentration and electron concentration. It is found that $m^*$ is 0.030 $\sim$ 0.171$m_0$ depending on the electron concentration. These data are all consistent with the effective mass of the conduction band electrons reported in heavily doped InAs (black and white data in Fig. S3), indicating that the electrons in (In,Fe)As reside in the conduction band, not in the hypothetical Fe-related impurity band with heavy effective mass.



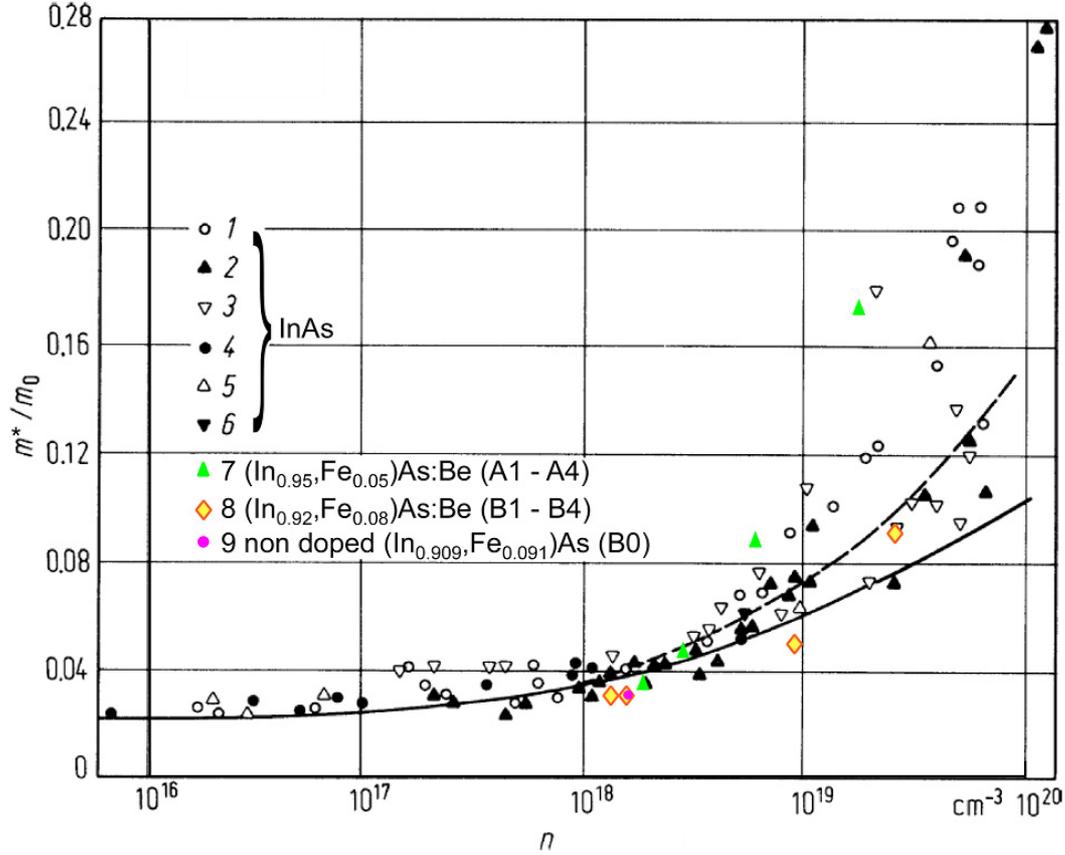

**Fig. S3.** Electron effective mass vs. electron concentration for various InAs-based semiconductors. Literature data (black and white) obtained by (1) the Seebeck effect, (2) infrared reflectivity, (3) magnetic susceptibility, (4) Faraday effect, (5) recombination radiation, (6) cyclotron resonance (after Ref. 2 and references therein). Our data (colored) are obtained by the Seebeck effect for (7) Be doped $(In_{0.95},Fe_{0.05})As$ samples (A1 - A4), (8) Be doped $(In_{0.92},Fe_{0.08})As$ samples (B1 - B4), and (9) a non-doped $(In_{0.909},Fe_{0.091})As$ sample (B0).

## Doped Be concentration vs. electron concentration

Figure S4 shows the doped Be concentration $n_{Be}$ vs. electron concentration $n$ for sample series A and B. $n$ are measured at room temperature. When doped at low growth temperature (236°C), Be atoms become donors rather than acceptors, whereas Be atoms become acceptors when doped at a normal growth temperature of InAs (400°C). Note that similar donor behavior of group II dopants in InAs grown at low substrate temperature has been observed for the case of Mn in (In,Mn)As [3]. The relationship between $n_{Be}$ vs. $n$ is not trivial. The highest $n$ is obtained when $n_{Be}$ is around $10^{19}$ cm$^{-3}$. At this doping level, the electron concentration is twice as large as the Be concentration. This suggests that Be



atoms reside at interstitial positions and act as double donors. The double donator behavior of interstitial Be has also been observed when doped in silicon carbide [4]. However, when increasing $n_{Be}$ up to $10^{20}$ cm$^{-3}$, $n$ decreased to $10^{18}$ cm$^{-3}$. Be doping levels except for $n_{Be} \sim 10^{19}$ cm$^{-3}$ resulted in $n < n_{Be}$.

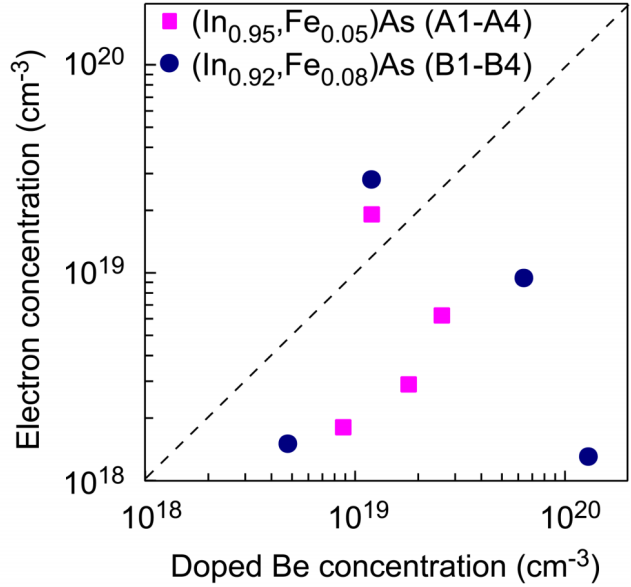

**Fig. S4.** Doped Be concentration vs. electron concentration.

**Magnetic Anisotropy of (In,Fe)As**

Fig. S5 shows the *M-H* curves of sample B4 measured with magnetic field *H* applied in the film plane and perpendicular to the plane. These two curves are parts of the *M-H* curves shown Fig. 4f and Fig. 5f of the main text for $H > 0.1$ Tesla. The *M-H* curve measured with *H* applied in the film plane saturates much faster than that with *H* applied perpendicular to the plane. The difference appearing for $H > 0.1$ Tesla comes from the shape anisotropy of the macroscopic matrix phase. Therefore, sample B4 includes macroscopic ferromagnetic matrix structures with size much larger than the film thickness of 100 nm. This fact is also consistent with the results of MCD measurements.



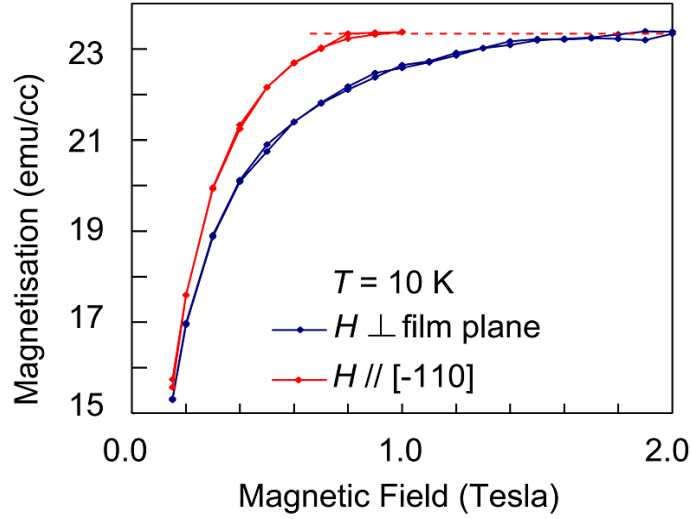

**Fig. S5.** *M-H* curves for $H > 0.1$ Tesla of sample B4 measured with a magnetic field *H* applied in the film plane (red) and perpendicular to the film plane (blue). These two curves are parts of Fig. 4f and Fig. 5f of the main text. It is clear that the in-plane *M-H* curve saturates much faster than the perpendicular-to-plane *M-H* curve, indicating that there is indeed magnetic shape anisotropy of the matrix phase.

To investigate the in-plane magnetic anisotropy of (In,Fe)As, we have studied the in-plane anisotropic magnetoresistance (AMR), and we clearly observed a two-fold anisotropy along the [-110] direction, and 8-fold symmetric anisotropy along the crystal axes of (In,Fe)As. This result supports the macroscopic intrinsic ferromagnetism of this new material.

To study AMR, we have fabricated a 10-nm thick n-type $(In_{0.94},Fe_{0.06})As$ layer doped with $8 \times 10^{18}$ cm$^{-3}$ electrons. The Curie temperature was estimated to be about 30 K from the Arrott plot of the MCD-*H* curves. Hall bars along the [110] and [-110] directions were fabricated for AMR measurements. The resistance of the Hall bars were measured by the 4-terminal method, with a magnetic field of 8.7 kG applied in-plane and rotate around the sample from $\alpha = 0$ to 360°, where $\alpha$ is the angle between the magnetic field and the [100] crystal axis. The AMR of a cubic crystal based on the symmetry argument is given by [5],

$\Delta\rho/\rho_{avr} = C_1\cos(2\phi) + C_{1,c}\cos(4\varphi - 2\phi) + C_2\cos(2\varphi) + C_4\cos(4\varphi).$ (S4)

Here $\Delta\rho = \rho - \rho_{avr}$, with $\rho_{avr}$ the averaged resistivity, $\phi$ is the angle between the magnetisation and the current, $\varphi = \alpha - \pi/4$ is the angle between the magnetisation and the [110] axis. The first term is the non-crystalline AMR, which originates from the s-d scattering effect. The third and the fourth terms are the crystalline AMR, which are related to the symmetry of the crystal. The higher order crystalline terms are usually neglected in



this equation (but see below). The second term is the crossed non-crystalline / crystalline terms. The first and the second terms in (In,Fe)As are found to be negligible because we observed no clear $\phi$-dependence of AMR for the [110] and [-110] Hall bars. This is quite reasonable because there is no d-state at the Fermi level, thus there is no s-d scattering. We further found that an 8-fold symmetric anisotropy term should be added to explain our experiment result.

$$\Delta\rho/\rho_{avr} = C_2\cos(2\varphi) + C_4\cos(4\varphi) + C_8\cos(8\varphi). \tag{S5}$$

Figure S6 shows $\alpha$ polar plot of $\Delta\rho/\rho_{avr}$ - $\Delta\rho_{min}/\rho_{avr}$ taken from a [110] Hall bar measured at 20 K. We clearly observed a two-fold symmetric anisotropy along the [-110] direction. There are additional AMR peaks at approximately 0, $\pi/4$, $\pi/2$, $\pi$, $5\pi/4$ and $3\pi/2$, revealing an 8-fold symmetric anisotropy term. The $C_2$, $C_4$ and $C_8$ coefficients can be uniquely obtained by solving

$$\begin{pmatrix} 1 & 1 & 1 \\ 0 & -1 & 1 \\ -1 & 1 & 1 \end{pmatrix}\begin{pmatrix} C_2 \\ C_4 \\ C_8 \end{pmatrix} = \begin{pmatrix} \Delta\rho(\varphi=0)/\rho_{avr} \\ \Delta\rho(\varphi=\pi/4)/\rho_{avr} \\ \Delta\rho(\varphi=\pi/2)/\rho_{avr} \end{pmatrix} \tag{S6}$$

The results are $C_2 = -0.038\%$, $C_4 = 0\%$, $C_8 = 0.025\%$. The calculated curve using these values in Eq. S5 is shown by the solid black curve, and explains reasonably well the experimental result (red squares). Note that such a crystalline AMR effect is a novel characteristic of single-crystal ferromagnetic materials. This effect has been observed only for the case of (Ga,Mn)As among FMSs. Since crystalline AMR reflects the symmetry of a macroscopic crystal, our AMR result strongly supports the intrinsic ferromagnetism in this macroscopic (In,Fe)As layer. Furthermore, it is well-known that the crystalline anisotropy of the magnetoresistance reflects the anisotropy of magnetization [5,6], thus this result also reveals 2-fold and 8-fold symmetric anisotropy of magnetisation in (In,Fe)As. The symmetry of AMR is related to the symmetry breaking due to macroscopic ferromagnetism, thus this cannot be explained by MR of nanoclusters or weak localization.



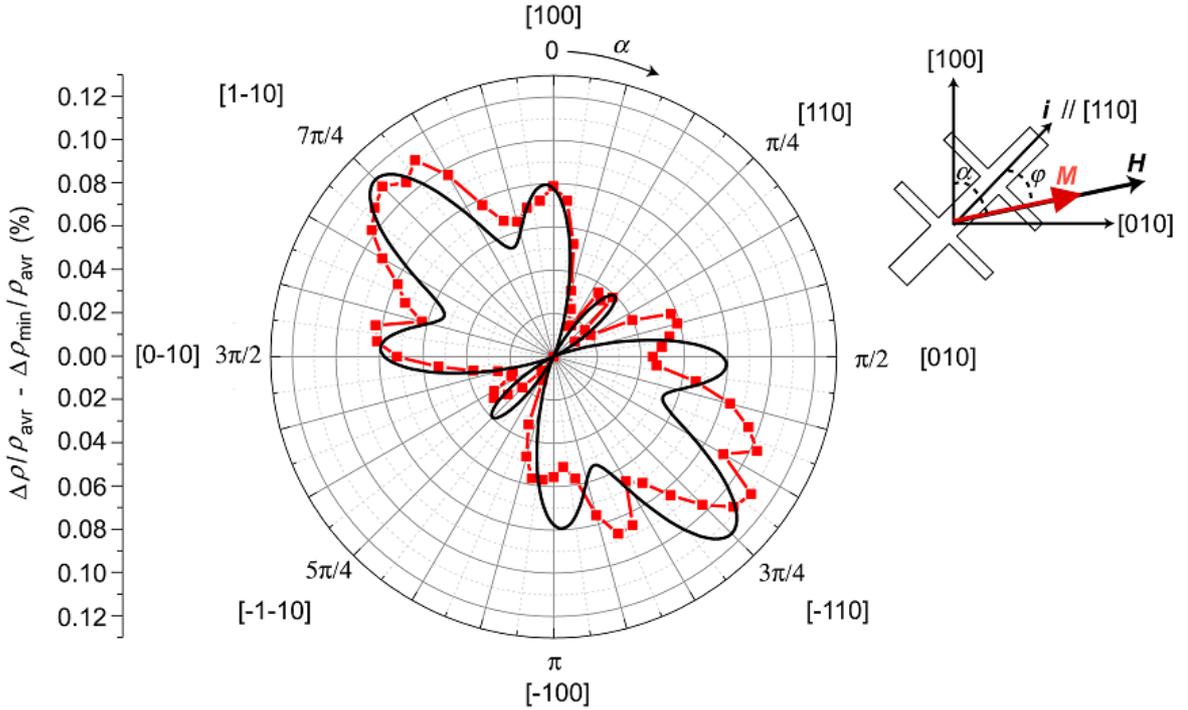

**Fig. S6.** $\alpha$ polar plot of the magnetoresistance ($\Delta\rho/\rho_{\mathrm{avr}}$ - $\Delta\rho_{\mathrm{min}}/\rho_{\mathrm{avr}}$) of a Hall bar along the [110] direction of a 10 nm-thick $(In_{0.94},Fe_{0.06})$As layer, measured at 20 K. The applied magnetic field $H$ was large enough (8.7 kG) so that the direction of magnetisation $M$ is the same as that of $H$. We clearly observed a two-fold anisotropy along the [-110] direction, and an 8-fold symmetric anisotropy along the crystal axes of (In,Fe)As, thus strongly supporting the macroscopic intrinsic ferromagnetism of this material. The red squares are experimental data. The solid black curve is calculated using $C_2$= −0.038%, $C_4$= 0%, $C_8$ = 0.025% in Eq. (S5).

**Origin of negative magnetoresistance**

There are several suggested mechanisms for the negative magnetoresistance shown in Fig. 5g of the paper. These includes the MR effect of granular systems [7], the MR effect observed in paramagnetic InAs:Mn [8], and the weak localization effect [9]. The MR effect of granular systems cannot explain the temperature dependence of MR shown in Fig. 5g of the main text. If the observed MR was due to the spin-dependent transport of electrons between ferromagnetic nanoclusters through the semiconducting InAs matrix, its order of magnitude would be given by $2P^2/(1-P^2)$, where $P$ is the spin-porälisation of electrons in the nanoclusters. While this may give finite negative MR ~ $M^2$ at $T < T_C$, no MR would be observed for $T > T_C$ because $P = 0$. Even if taking into account the Pauli paramagnetism of electrons in the nanoclusters, the spin-polarisation is only of the order of ($\mu_B\mu_0H / E_F$)~$10^{-3}$ at for 1 Tesla, giving MR ~ $10^{-6}$ that is negligible. In contrast, in sample



A4, the MR was -0.55 % at 10 K, and even at 70 K (> $T_C$ = 34 K for this sample) MR is still - 0.23%, the same order of magnitude.

The negative MR observed in homogeneous n-type paramagnetic InAs:Mn (ref. 8) does not contradict our results, but it supports the spin-disorder scattering mechanism in our (In,Fe)As at $T > T_C$. With the spin-disorder scattering mechanism, it is natural to explain our result MR ~ $M^2$ even when the sample is paramagnetic (i.e. at $T > T_C$) (see, for example, ref. 10).

The weak localization mechanism can also be excluded, because it gives MR ~ $B^{1/2}$ (Refs. 9,11), which is different from MR ~ $M^2$ observed experimentally in our (In,Fe)As.

Finally, we have studied the in-plane anisotropic magnetoresistance (AMR), and we clearly observed the two-fold anisotropy along the [-110] direction, and 8-fold symmetric anisotropy along the crystal axes of (In,Fe)As. The symmetry of AMR is related to the symmetry breaking due to macroscopic ferromagnetism, thus this cannot be explained by MR of nanoclusters or weak localization.